# Influence of the Ion Coordination Number on Cation Exchange Reactions with Copper Telluride Nanocrystals

Renyong Tu[1], Yi Xie[1,2], Giovanni Bertoni[1,3], Aidin Lak[4], Roberto Gaspari[5], Arnaldo Rapallo[6], Andrea Cavalli[5,7], Luca De Trizio*[1] and Liberato Manna*[1]

[1] Department of Nanochemistry, Istituto Italiano di Tecnologia (IIT), via Morego, 30, 16163 Genova, Italy

[2] State Key Laboratory of Silicate Materials for Architectures, Wuhan University of Technology (WUT), No. 122, Luoshi Road, Wuhan 430070, P. R. China.

[3] IMEM-CNR, Parco Area delle Scienze, 37/A, 43124 Parma, Italy

[4] Drug Discovery and Development Istituto Italiano di Tecnologia (IIT), via Morego, 30, 16163 Genova, Italy

[5] CompuNet, Istituto Italiano di Tecnologia (IIT), via Morego, 30, 16163 Genova, Italy

[6] ISMAC – Istituto per lo Studio delle Macromolecole del CNR, via E. Bassini 15, 20133 Milano, Italy

[7] Department of Pharmacy and Biotechnology, University of Bologna, via Belmeloro 6, I-40126 Bologna, Italy

**ABSTRACT:** $Cu_{2-x}Te$ nanocubes were used as starting seeds to access metal telluride nanocrystals by cation exchanges at room temperature. The coordination number of the entering cations was found to play an important role in dictating the reaction pathways. The exchanges with tetrahedrally coordinated cations (*i.e.* with coordination number 4), such as $Cd^{2+}$ or $Hg^{2+}$, yielded monocrystalline CdTe or HgTe nanocrystals with $Cu_{2-x}Te/CdTe$ or $Cu_{2-x}Te/HgTe$ Janus-like heterostructures as intermediates. The formation of Janus-like architectures was attributed to the high diffusion rate of the relatively small tetrahedrally coordinated cations, which could rapidly diffuse in the $Cu_{2-x}Te$ NCs and nucleate the CdTe (or HgTe) phase in a preferred region of the host structure. Also, with both $Cd^{2+}$ and $Hg^{2+}$ ions the exchange led to wurtzite CdTe and HgTe phases rather than the more stable zinc-blende ones, indicating that the anion framework of the starting $Cu_{2-x}Te$ particles could be more easily deformed to match the anion framework of the metastable wurtzite structures. As hexagonal HgTe had never been reported to date, this represents another case of metastable new phases that can only be accessed by cation exchange. On the other hand, the exchanges involving octahedrally coordinated ions (*i.e.* with coordination number 6), such as $Pb^{2+}$ or $Sn^{2+}$, yielded rock-salt polycrystalline PbTe or SnTe nanocrystals with $Cu_{2-x}Te@PbTe$ or $Cu_{2-x}Te@SnTe$ core@shell architectures at the early stages of the exchange process. In this case, the octahedrally coordinated ions are probably too large to diffuse easily through the $Cu_{2-x}Te$ structure: their limited diffusion rate restricts their initial reaction to the surface of the nanocrystals, where cation exchange is initiated unselectively, leading to core@shell architectures. Interestingly, these heterostructures were found to be metastable as they evolved to stable Janus-like architectures if annealed at 200 °C under vacuum.

## INTRODUCTION

During the last decade cation exchange (CE) reactions have become one of the most important post-synthetic tools for the chemical transformation of nanomaterials.[1-8] Such reactions allow for the selective replacement of all (total CE) or a part (partial CE) of the cations of preformed ionic nanocrystals (NCs) with new desired guest cations, while retaining their size, shape and anion framework. This technique has been successfully applied to many ionic NCs, above all to those belonging to the II-VI, I-III-VI and IV-VI classes of semiconductors, *i.e.* to metal chalcogenides. Although the knowledge of the thermodynamics behind CE has been gradually consolidated, emerging studies are still aiming at elucidating the mechanisms involved in these reactions.[9-24] Depending on the specific transformation under analysis, the ingoing and outgoing cations can diffuse through vacancies (mainly cation vacancies) and/or interstitial lattice sites.[10-14,23] In principle, the morphology resulting from a partial CE experiment can be predicted and tailored as it is intimately connected to the mechanisms and the kinetics of ionic diffusion and replacement in the system under investigation. For example, the partial Cd-for-Pb exchange in PbX (X=S, Se, Te) NCs leads to the formation core@shell PbX@CdX heterostructures.[9,13,25-32] As demonstrated by Grodzinska *et al.*, these structures are metastable (*i.e.* kinetically driven), and evolve into Janus-like systems upon annealing at temperatures as low as 150 °C under vacuum.[33] These findings suggest that PbX@CdX NCs form as a result of the slow out-diffusion and replacement of Pb²⁺ ions that allow, statistically, for the unselective nucleation of the product zinc-blende (zb) CdX material on the whole surface of the starting rock-salt (rs) PbX NCs. This is also promoted by the absence of a preferential interface between zb-CdX and rs-PbX materials, as both have a cubic crystal structure, with almost no lattice mismatch between them.[13,27,29] Surprisingly, the inverse CE reaction, that is between CdX (X=S, Se) NCs and Pb²⁺ ions, results in segmented CdX/PbX heterojunctions.[34,35] These heterostructures, unlike the core@shell systems, are thermodynamically stable, evidencing that the two CE processes are characterized by different reaction pathways.

While a large selection of metal sulfides and selenides NCs are now accessible by CE reactions, few works have been reported instead on metal telluride NCs.[36] A consolidated CE procedure developed for MX (X=S, Se) NCs, in fact, might not necessarily work on the corresponding MTe NCs. This can be explained by considering that the M-Te bonds are weaker than the corresponding M-S and M-Se bonds, which makes the whole tellurium sublattice less rigid and thus more prone to deformation or dissolution during the same CE reaction. In this work, with the aim of expanding the library of metal telluride nanostructures accessible *via* CE, we studied exchange reactions involving cubic-shaped Cu₂₋ₓTe NCs and guest cations which are characterized by different coordination numbers in the host lattice: four for Cd²⁺ and Hg²⁺ ions (tetrahedral coordination) and six for Pb²⁺ and Sn²⁺ cations (octahedral coordination). It is important to underline that all the selected cations cannot form alloyed structures with the host material, therefore partial CE experiments necessarily lead to heterostructures. In analogy to the cases of Cu₂₋ₓS and Cu₂₋ₓSe NCs, CE reactions in Cu₂₋ₓTe NCs are promoted by tri-n-octyl phosphine (TOP, as soft Lewis base) and they are facilitated by the high density of copper vacancies. This allows for an efficient CE yield even at low temperature (*i.e.* room temperature) which is essential for the preservation of the size and morphology of the starting metal telluride NCs. The use of TOP, in some cases, was however not sufficient to effectively guarantee the CE transformation and the addition of N,N-dimethylethylenediamine (NND) was found to help in the effective stabilization of copper-ligand complexes and, thus, in shifting the equilibrium of the CE reaction towards the product materials.

**Scheme 1. CE reactions between cubic-shaped Cu₂₋ₓTe NCs and tetrahedrally (Cd²⁺, Hg²⁺) or octahedrally (Pb²⁺, Sn²⁺) coordinated cations, with coordination numbers (CN) 4 and 6, respectively.**

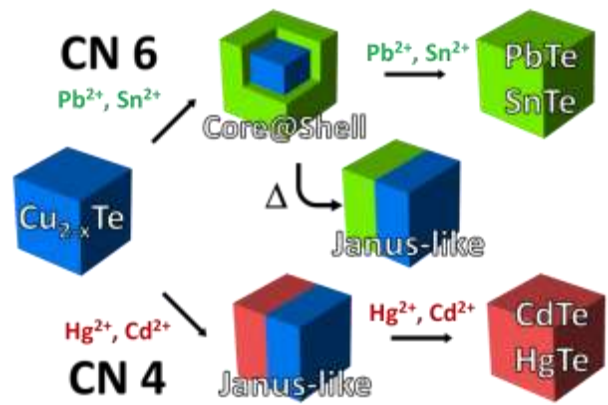

We show here that the coordination number of the guest cations drastically influences the kinetics of a CE reaction, with a direct consequence on the morphology of the resulting intermediate heterostructures. Cd²⁺ and Hg²⁺ ions, which adopt a tetrahedral coordination, are fast diffusers and, in total CE reactions, they lead to the formation of monocrystalline hexagonal (hex) CdTe and HgTe NCs, respectively. Partial CE experiments with such cations result in Janus-like NCs having sharp epitaxial CdTe/Cu₂₋ₓTe and HgTe/Cu₂₋ₓTe interfaces. It is noteworthy to underline that the hexagonal HgTe phase observed here is not known in the bulk, and therefore our reaction scheme further expands the set of metastable materials accessible *via* CE.[37-40] On the other hand, CE reactions involving Pb²⁺ or Sn²⁺ cations, which are comparatively slower diffusers as they adopt an octahedral coordination, start with the formation of a polycrystalline rs-PbTe or rs-SnTe shell around the parent Cu₂₋ₓTe NCs. The exchange proceeds then further, with the total replacement of Cu⁺ ions and the formation of polycrystalline rs-PbTe or rs-SnTe NCs. Both Cu₂₋ₓTe@PbTe and Cu₂₋ₓTe@SnTe core@shell NCs are metastable (kinetically accessed) as they evolve into Janus-like architectures upon annealing at 200 °C under vacuum. The products of our transformations are depicted in scheme 1. The results of our work suggest that the coordination number of the guest cations has a profound influence on the overall kinetics of CE reactions. Tetrahedrally coordinated cations, being fast diffusers, are able to initiate the CE in a desired site on the surface of the starting Cu₂₋ₓTe NCs. The exchange process, then, proceeds from there at the expenses of the host material forming NCs composed of two distinct domains sharing a flat interface, that is Janus-like heterostructures. Octahedrally coordinated cations, on the other hand, can rapidly engage in CE on the surface of host NCs, but then cannot quickly diffuse toward the core of the host NCs. This leads to the formation of kinetically accessed core@shell NCs with tunable shell thickness.

EXPERIMENTAL SECTION

**Materials.** Copper (II) acetylacetonate (Cu(acac)₂, 97%), copper (I) chloride (CuCl, 99.99%), oleylamine (OLAM, >70%), octadecene (ODE, 90%), tri-n-octylphosphine (TOP, 97%), trioctylphosphine oxide

(TOPO, 99%), lithium bis(trimethylsilyl)amide (LiN(SiMe$_3$)$_2$, 97%), N,N-dimethylethylenediamine (NND, ≥98.0% ), mercury (II) chloride (HgCl$_2$, ≥99.5% ), cadmium iodide (CdI$_2$, 99% ), lead (II) acetylacetonate (Pb(acac)$_2$, ≥95%) and tin (II) chloride (SnCl$_2$, 98%), were purchased from Sigma-Aldrich, tellurium powder (99.999%) and selenium powder (99.99%) from Strem Chemicals, ethanol (anhydrous, ≥99.8%), methanol (anhydrous, ≥99.8%), toluene (anhydrous, ≥99%), and chloroform (anhydrous, ≥99%) from Carlo Erba reagents. All chemicals were used as received without further purification and all reactions were carried out under nitrogen using standard air-free techniques.

**Synthesis of Cu$_{2-x}$Te NCs.** The synthesis of cubic-shaped Cu$_{2-x}$Te NCs was carried out following the work of Li *et al.* with minor modifications.[41] In a typical reaction, a Te precursor mixture was prepared by mixing 1.25 mL of a 2 M TOP-Te solution with 5 mL of a 0.5 M ODE-LiN(SiMe$_3$)$_2$ solution. The TOP-Te solution was obtained by dissolving tellurium powder in TOP at 150 °C for two hours, while a clear ODE-LiN(SiMe$_3$)$_2$ solution could be prepared at room temperature (RT) by sonication for 10 min. A mixture of 0.655 g of Cu(acac)$_2$ (2.5 mmol), 3.866 g of TOPO (10 mmol) and OLAM (50 mL) was degassed in a 250 mL 3-necks flask at 100 °C for 30 min. The temperature was then raised to 160 °C and 1.5 mL of TOP were added to the flask. Eventually, the as-prepared Te precursor mixture was rapidly injected into the reaction flask and then the temperature was set to 220 °C and the reaction was allowed to proceed for 30 min at that temperature. The flask was rapidly cooled to RT with a water bath to quench the reaction and 10 mL of chloroform were added to the reaction flask when the temperature dropped to about 70 °C. The NCs were isolated by centrifugation and washed twice by re-dissolution in chloroform and precipitation with the addition of ethanol. Eventually the NCs were dispersed in chloroform and stored in a N$_2$ filled glove-box.

**CE reactions with Cd$^{2+}$ ions.** In a typical CE reaction, 0.1 mL of TOP and 0.1 mL of NND were added to a glass vial containing a 4 mL dispersion of Cu$_{2-x}$Te NCs (0.02 mmol of Cu$^+$ ions) in chloroform. Afterward a desired amount of a 0.1 M solution of CdI$_2$ in ethanol was added to the vial and the resulting mixture was stirred for 30 min at RT (see Table S1 for further details). The resulting NCs were washed twice by precipitation with addition of ethanol followed by re-dissolution into chloroform and were eventually stored in the glove-box.

**CE reactions with Hg$^{2+}$ ions.** 0.1 mL of TOP were added to a glass vial containing a dispersion of Cu$_{2-x}$Te NCs (0.02 mmol of Cu$^+$ ions) in 4 mL of chloroform. A desired amount of a 0.1 M solution of HgCl$_2$ in ethanol was then added to the vial at RT (see Table S1 for further details). The reaction mixture was stirred for 30 min and the resulting NCs were washed twice by precipitation with addition of ethanol followed by re-dissolution into chloroform and eventually stored in the glove-box.

**CE reactions with Pb$^{2+}$ ions.** 0.1 mL of TOP and 0.1 mL of NND were added to a glass vial containing a dispersion of Cu$_{2-x}$Te NCs (0.02 mmol of Cu$^+$ ions) in 4 mL of toluene. Afterward a desired amount of a 0.1 M solution of Pb(acac)$_2$ in chloroform was added to the vial and the resulting mixture was stirred for 30 min at RT (see Table S1 for further details). The resulting NCs were washed twice by precipitation with addition of ethanol followed by re-dissolution into toluene and eventually stored in the glove-box.

**CE reactions with Sn$^{2+}$ ions.** In a typical CE reaction, 0.1 mL of TOP were added to a dispersion of Cu$_{2-x}$Te NCs (0.02 mmol of Cu$^+$ ions) in 4 mL of toluene in a glass vial. Afterward a desired amount of a 0.1 M solution of SnCl$_2$ in ethanol was added to the vial at RT under stirring (see Table S1 for further details). The reaction was stopped after 30 min. The resulting NCs were washed twice by precipitation with addition of ethanol followed by re-dissolution into toluene and eventually stored in the glove-box.

**TEM measurements.** The samples were prepared by dropping dilute solutions of NCs onto carbon coated gold grids, which were then placed under vacuum to preserve them from oxidation. Low-resolution transmission electron microscopy (TEM) measurements were carried out on a JEOL JEM-1100 transmission electron microscope operating at an acceleration voltage of 100 kV. High Resolution TEM (HRTEM) was performed in a JEOL JEM-2200FS microscope equipped with a 200 kV field emission gun, a CEOS spherical aberration corrector in the objective lens, enabling a spatial resolution of 0.9 Å, and an in column energy filter. High angle annular dark field images were acquired on the same microscope in scanning mode (STEM) with a nominal probe size of 0.2 nm, and an inner cut-off angle of the annular detector of 75 mrad.

**X-Ray Diffraction (XRD) measurements.** The XRD analysis was performed on a PANalytical Empyrean X-ray diffractometer equipped with a 1.8 kW CuKα ceramic X-ray tube, PIXcel$^{3D}$ 2x2 area detector and operating at 45 kV and 40 mA. Specimens for the XRD measurements were prepared in a glove box by dropping a concentrated NCs solution onto a quartz zero-diffraction single crystal substrate. The diffraction patterns were collected at ambient conditions using a parallel beam geometry and symmetric reflection mode. XRD data analysis was carried out using the HighScore 4.1 software from PANalytical. The cell refinement of the pseudo-cubic Cu$_{2-x}$Te structure was performed using the house-made Pindex[42] program in a search region 7.3Å<a<7.7Å, 7.3Å<b<7.7Å, 7.2Å<c<7.6Å, *i.e.* covering the range of uncertainty of the STEM analysis. The Le Bail fit of the experimental XRD pattern was performed using the Fox program.[43-44] The Rietveld crystal structure analysis was performed using FullProf Suite program.

**Elemental Analysis.** This was carried out *via* Inductively Coupled Plasma Atomic Emission Spectroscopy (ICP-AES), using an iCAP 6500 Thermo spectrometer. Samples were dissolved in HCl/HNO$_3$ 3:1 (v/v). All chemi-cal analyses performed by ICP-AES were affected by a systematic error of about 5%.

## RESULTS AND DISCUSSION

In order to investigate CE reactions in metal tellurides, cubic-shaped $Cu_{2-x}Te$ NCs were employed as starting seeds. The choice of this material relies on two main advantages: 1) it is well known that $Cu^+$ ions can be easily extracted from copper chalcogenide NCs in the presence of alkyl phosphines and replaced with stronger Lewis acids;[36] 2) sub-stoichiometric copper chalcogenide NCs are characterized by a high density of Cu vacancies which favor the cation diffusion.[11,15,38] Thanks to these peculiar properties, CE reactions in such material are expected to take place already at low temperature, and this circumvents the tedious problems that are often connected to high temperature routes (etching and/or deformation of the host NCs). The starting $Cu_{2-x}Te$ NCs, synthesized following a procedure developed by Li et al.,[41] have a $Cu_{1.55}Te$ stoichiometry, as measured via ICP elemental analysis, and exhibit a good size dispersion with a mean size of 20.1±2.1 nm, as shown in Figure 1a and Figure S1 of the Supporting Information (SI).

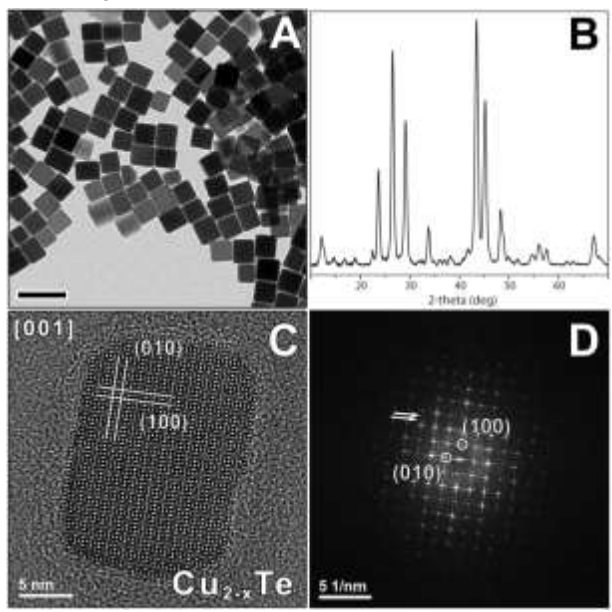

**Figure 1.** (a) Low resolution TEM image of as-synthesized $Cu_{2-x}Te$ NCs. The scale bar is 50 nm. (b) XRD pattern obtained from dropcast solutions of $Cu_{2-x}Te$ NCs. (c) High resolution TEM (HRTEM) image of a $Cu_{2-x}Te$ NC with the corresponding (d) Fast Fourier Transform (FFT). The FFT pattern shows a pseudo-cubic structure with $a$ and $b$ close to 7.5 Å. The arrows mark the spots in the FFT corresponding to the 3× superstructure, here clearly visible in the [100] direction.

As already pointed out by Li et al.,[41] the crystal structure of such NCs does not match with any known $Cu_{2-x}Te$ bulk phase, but we found that it closely resembles a pseudo-cubic structure, with lattice length $a$=7.22 Å, first reported by Thompson in 1949.[45] Our $Cu_{2-x}Te$ NCs, as evidenced by HRTEM images (see Figure 1c-d), show, indeed, a pseudo-cubic structure, which can be better represented by an orthorhombic phase with $a$=7.50 Å, $b$=7.53 Å and $c$=7.48 Å. Both FFT and HRTEM images highlight also the presence of a superstructure in which the actual unit cell of the pseudo-cubic structure (1×1×1 structure) repeats with 3, 3 and 4 times modulations (3×3×4 structure) along the $a$, $b$ and $c$ directions, respectively. Indeed, the 1×1×1 pseudo-cubic structure is compatible with the most intense peaks of the pattern, but it does not reproduce the minor reflections that, on the other hand, can be explained by taking into account a 3×3×4 superstructure (see SI for additional details).

In our sets of experiments, the $Cu_{2-x}Te$ NCs were exposed to divalent cations commonly used in CE reactions, such as $Cd^{2+}$, $Hg^{2+}$, $Pb^{2+}$ and $Sn^{2+}$. $Cd^{2+}$ and $Hg^{2+}$ ions, adopting a tetrahedral coordination, have relatively small ionic radii (78 pm and 96 pm, respectively),[46] not far from that of the $Cu^+$ cations (60 pm with coordination 4 and 77 pm with coordination 6) of the $Cu_{2-x}Te$ NCs, which should allow them to rapidly diffuse via vacancies or interstitials inside the "host" NCs. $Pb^{2+}$ and $Sn^{2+}$ cations, which adopt an octahedral coordination with $Te^{2-}$ anions, have on the other hand much larger ionic radii (119 pm and 118 pm, respectively)[46] which, most likely, would limit their diffusion through vacant cation sites only and not through interstitial sites.[14,47] The final and intermediate nanostructures, formed by exposing the $Cu_{2-x}Te$ NCs to such cations at RT, were investigated in order to get insights over the possible different kinetics of CE.

Before going into details on the analysis of the product structures, we would like to first discuss the role of the Lewis bases used in our CE reactions. Although in all the cases discussed here the presence of TOP was essential for the CE transformation to take place, the complete $Cu_{2-x}Te \rightarrow CdTe$ and $Cu_{2-x}Te \rightarrow PbTe$ transformations were made possible only by the additional use of NND. In these cases, the solvation energy contribution provided by TOP alone was, most likely, not high enough to achieve complete CE at RT, since only partial CE was observed even at longer reaction times (overnight). On the other hand, NND is not known to strongly bind to $Cu^+$ ions, and, indeed, when CE experiments were performed using NND in the absence of TOP, the exchange took place only to a very small extent. In organometallic chemistry NND is well known to form very stable thermochromic $[Cu(NND)_2]X_2$ (X=BF$_4^-$, ClO$_4^-$, NO$_3^-$) complexes.[48-50] We believe that this bidentate molecule is able to promote CE reactions by increasing the stability of $Cu^+$-TOP complexes and, thus, favoring the extraction of copper cations.[36]

We start by discussing CE experiments with tetrahedrally coordinated ions. As reported in the low resolution TEM images of Figure S2a-b of the SI, the shape and size distribution of the starting cubic-shaped $Cu_{2-x}Te$ NCs was retained after CE. Figure 2 summarizes HRTEM and XRD analyses performed on CdTe and HgTe NCs resulting from total CE experiments with $Cd^{2+}$ and $Hg^{2+}$ ions, respectively. The ICP elemental analysis confirmed the complete exchange reaction for both $Cd^{2+}$ and $Hg^{2+}$ CE transformations, with a residual amount of copper being lower than 2% (see Table S1 for details). In the exchange reaction with $Cd^{2+}$ cations the XRD pattern of the obtained NCs was consistent with bulk hexagonal (hex) CdTe with $a,b$=4.58 Å and $c$=7.52 Å (ICSD number 98-062-0518, see Figure 2c). HRTEM analyses performed on Cd-exchanged $Cu_{2-x}Te$ NCs evidenced their defect-free and monocrystalline habit, confirming, at the same time, their hexagonal crystal structure, as found by XRD (see Figure 2a). A comparison between the hex-CdTe and the cubic $Cu_{2-x}Te$ structures suggests that a slight distortion of the anion sublattice took place during the CE reaction. This supports, once again, that the $Te^{2-}$ anion sublattice in metal tellurides is prone to deformation even at mild reaction conditions. Another interesting feature of this transformation is that an hex-CdTe phase is observed upon CE rather than the more stable cubic CdTe phase as, most likely, the transition to the former requires a lower activation barrier.[51] In a control experiment, we observed indeed that our hex-CdTe NCs, if annealed at ~200 °C under inert atmosphere, undergo a phase transformation to the more stable cubic cadmium telluride (a= 6.47Å, ICSD 98-062-0531, see Figure S4 of the SI).

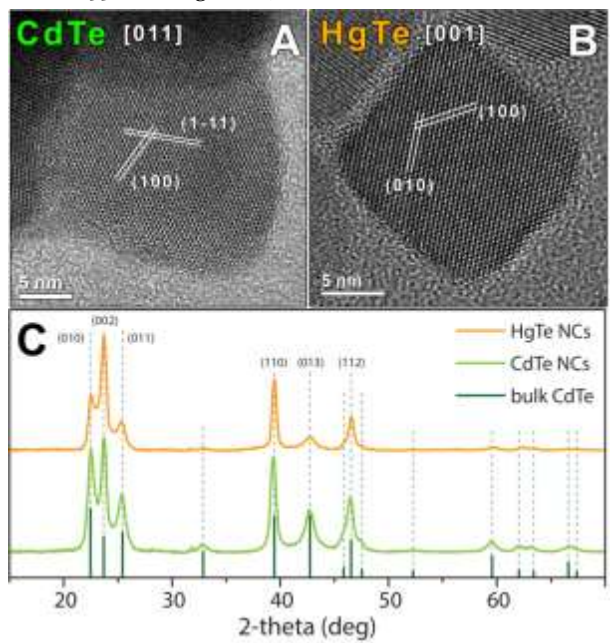

**Figure 2.** HRTEM images of a CdTe (a) and a HgTe (b) NCs obtained *via* total CE of $Cu_{2-x}Te$ NCs with $Cd^{2+}$ and $Hg^{2+}$ cations, respectively. CdTe and HgTe NCs, as evidenced by HRTEM analysis, are monocrystalline. (c) XRD patterns obtained from dropcast solutions of CdTe (green pattern) and HgTe (orange pattern) NCs with the corresponding bulk reflections of CdTe (dark green, ICSD number 98-062-0518).

These results indicate that, when CE is performed at RT, the hexagonal phase is kinetically accessed. The preferential transformation from the pseudo-cubic $Cu_{2-x}Te$ phase to the high energy hex-CdTe can be explained by considering the anionic sublattices of $Cu_{2-x}Te$ and CdTe structures. For this scope, we exploit the fact that the contrast in HAADF imaging is dominated by the heavier Te atoms in $Cu_{2-x}Te$ (see Figure 3). The Te anions of the hex-CdTe structure can indeed be superimposed, in the [100] projection, to the brightest spots of the $Cu_{2-x}Te$ HAADF image. These spots form a zig-zag ABAB… pattern (see the sketch in Figure 3) which resembles the typical stacking in the wurtzite structure. Moreover, the unit cell of CdTe projected along [100] has nearly the same periodicity (7.5 Å × 7.9 Å) of the $Cu_{2-x}Te$ pseudo-cubic faces. These considerations suggest that the transition from the pseudo-cubic $Cu_{2-x}Te$ to the hexagonal CdTe phase can be initiated on the facets of the $Cu_{2-x}Te$ cubes with reduced distortion of the anionic lattice. On the other hand, the low-index projection of the CdTe cubic phase ($a$=6.47 Å) poorly matches the periodicity of the $Cu_{2-x}Te$ pseudo-cubic faces, thereby suggesting that a direct transition from $Cu_{2-x}Te$ to cubic CdTe would require a larger structural rearrangement and thus a higher activation energy.

In analogy with the case of $Cd^{2+}$ CE reactions, the exposure of $Cu_{2-x}Te$ NCs to $Hg^{2+}$ ions resulted in monocrystalline HgTe NCs, again with a hexagonal crystal structure, as confirmed by both XRD and HRTEM analyses (see Figure 2b-c and Table S1). Although no match to any known hexagonal HgTe phase was found, the XRD pattern of HgTe NCs could be reconstructed using the hexagonal structural model of CdTe phase with $a$, $b$ = 4.55 Å and $c$ = 7.49 Å. These findings suggest that, under our synthetic conditions and as already observed in other works in the literature, it was possible to access a metastable crystal phase by means of a CE reaction.[37-40] A control experiment confirmed the low stability of such hexagonal phase, which transformed into the stable cubic HgTe phase upon annealing at 160 °C (see Figure S5 of the SI).

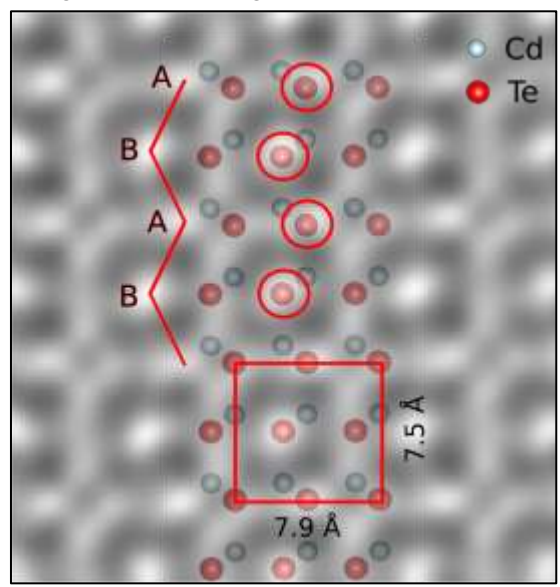

**Figure 3.** HAADF image from the pristine pseudo-cubic $Cu_{2-x}Te$. The bright dots are related to the heavier atoms (Te) in the structure. The hexagonal CdTe structure, projected along the [100] direction, is superimposed, showing a good match of the Te atoms (red) with the pristine anion sublattice. As it can be seen in the sketch, the "host" anion sublattice tends to favor the ABAB… layer stacking, promoting the formation of the hexagonal CdTe structure.

Since the product CdTe and HgTe structures are extremely similar, it is plausible that in both Cu$_{2-x}$Te→CdTe and Cu$_{2-x}$Te→HgTe transformations the same mechanism is involved. In order to study the evolution of CE reactions with such cations, partial exchange experiments were performed and the corresponding heterostructures were analyzed. As it is possible to estimate from the low resolution TEM images reported in the SI, in both Cd$^{2+}$ and Hg$^{2+}$ partial CE reactions the size and the shape of the starting Cu$_{2-x}$Te NCs were preserved (see Figure S2c-d and Table S1 of the SI). The XRD analysis of the product NCs, shown in figure 4c, evidenced the presence of two crystal phases: the starting pseudo-cubic Cu$_{2-x}$Te phase and the hexagonal CdTe or HgTe phases. The occurrence of two distinct phases was supported by the HRTEM characterization which clearly showed that, upon partial CE, Janus-like heterostructures, consisting of a CdTe (or HgTe) and a Cu$_{2-x}$Te domains, were formed (see Figure 4a,b). As regarding the epitaxial relations between the two domains, the CdTe (or HgTe) domain forms a relatively straight interface due to the good matching between the (100) planes of Cu$_{2-x}$Te and the (002) planes of HgTe (or CdTe) (see Figure 4b). However, a similar matching cannot be obtained on the orthogonal plane, so that a slight tilt and mismatch exist between the (001) planes of Cu$_{2-x}$Te and the (110) planes of CdTe (or HgTe), forming an oblique and probably strained interface (see Figure 4a).

er comparison, the experimental reflections measured for the starting Cu$_{2-x}$Te NCs (see Figure 1) are also reported by means of grey bars.

It is known that Janus-like or striped nano-architectures formed upon CE are, in general, thermodynamically driven, as they minimize the overall interfacial energy.[36] We believe that in our Cd$^{2+}$ and Hg$^{2+}$ CE reactions the relatively small four-coordinated ions, favored by the presence of a high density of Cu-vacancies, are able to rapidly access and diffuse through the host NC and to nucleate the new phase at a preferred site (most likely at one of the eight vertices of the starting nanocubes).[11,15,38] The nucleation of CdTe or HgTe phases occurs usually at one site per NC and the exchange proceeds from there with the formation of a single low energy interface between the CdTe or HgTe and the Cu$_{2-x}$Te domains in order to keep the overall energy to a minimum value. The nucleation of additional HgTe or CdTe domains is energetically disadvantaged and kinetically has to compete with the high diffusion rate of Hg$^{2+}$ or Cd$^{2+}$ ions, which participate to the growth of the existing HgTe or CdTe domain. Nevertheless, this process is only statistically less probable. A detailed HRTEM analysis of Cu$_{2-x}$Te/CdTe heterostructures and CdTe NCs evidenced, indeed, that in a few cases a "simultaneous" formation and growth of different CdTe domains in a single host NC occurred (see Figure S6 of the SI).

We consider now the products observed when working with octahedrally coordinated cations. As found for tetrahedrally coordinated ions, in the Cu$_{2-x}$Te→PbTe and Cu$_{2-x}$Te→SnTe transformations the size and the shape of the starting copper telluride NCs were retained, together with a complete replacement of the original Cu$^+$ cations, as confirmed by ICP analysis (see Figure S3a-b and Table S1 of the SI). The XRD characterization of the product NCs confirmed their crystallinity, with the formation of rs-PbTe phase (also called altaite) in the CE with Pb$^{2+}$ ions and rs-SnTe when working with Sn$^{2+}$ (see Figure 5c). In both cases, however, as it is possible to notice from both low and high resolution TEM images, the vast majority of the resulting PbTe and SnTe NCs was polycrystalline (see Figure 5a-b and Figure S3a-b of the SI). HRTEM images evidence that each PbTe and SnTe NC is composed of multiple domains, even though the original Cu$_{2-x}$Te NC morphology was fully retained. This suggests that in each NC the exchange started from different nucleation sites and from there it proceeded up to the total replacement of copper ions.

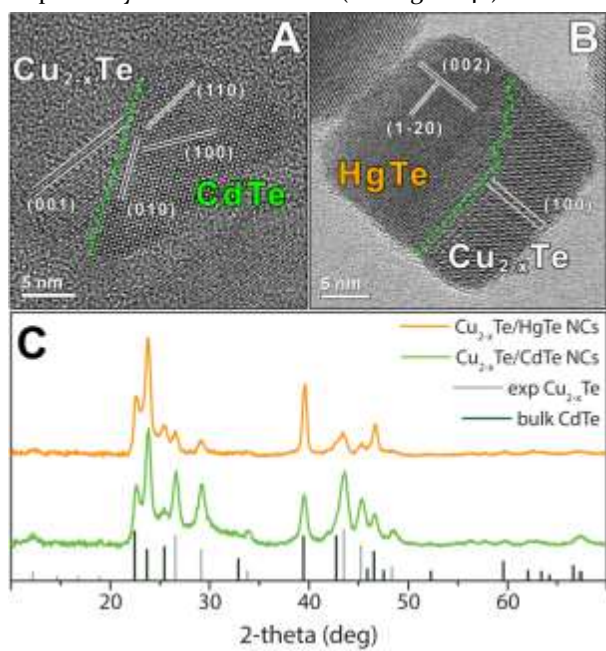

**Figure 4.** HRTEM images of Cu$_{2-x}$Te/CdTe (a) and Cu$_{2-x}$Te/HgTe (b) heterostructures prepared by partial CE of Cu$_{2-x}$Te NCs with Cd$^{2+}$ and Hg$^{2+}$ cations, respectively. As it is possible to appreciate from HRTEM images, both Janus-like NCs are made of a monocrystalline Cu$_{2-x}$Te domain and a monocrystalline hex-CdTe or hex-HgTe domain, with a sharp interface between the two domains indicated by a green dashed line. (c) XRD patterns obtained from dropcast solutions of Cu$_{2-x}$Te/CdTe NCs (green pattern) and Cu$_{2-x}$Te/HgTe (orange pattern) NCs with the corresponding bulk reflections of CdTe (dark green, ICSD number 98-062-0518). For a clear-

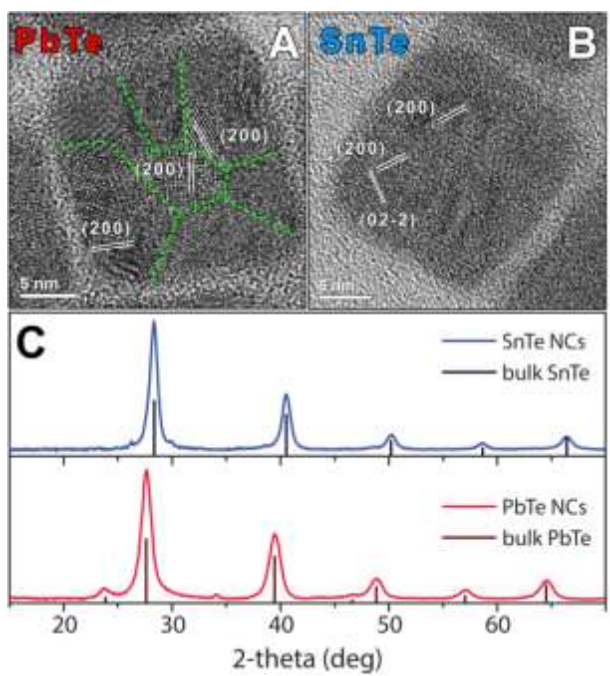

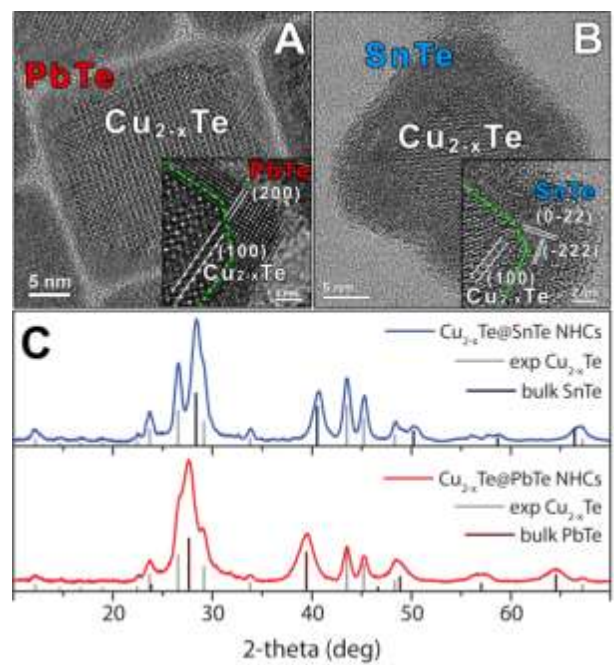

**Figure 5.** HRTEM images of a PbTe (a) and a SnTe (b) NCs obtained *via* total CE of $Cu_{2-x}Te$ NCs with $Pb^{2+}$ and $Sn^{2+}$ cations, respectively. Both PbTe and SnTe NCs are polycrystalline as highlighted in (a) with green dashed lines delimiting the various domains. (c) XRD patterns obtained from dropcast solutions of PbTe (red pattern) and SnTe (blue pattern) NCs with the corresponding bulk reflections: SnTe (dark blue bars, ICSD number 98-065-2759) and altaite PbTe (dark red bars, ICSD number 98-006-3099).

Partial CE reactions were performed in order to get insights on the nucleation and the growth processes in CE transformations with octahedrally coordinated cations. In both $Sn^{2+}$ and $Pb^{2+}$ CE experiments, heterostructures with a core@shell morphology were observed, as it could be seen from low and high resolution TEM images (see Figure 6a-b and Figure S3c-d of the SI). XRD and ICP analyses confirmed that the product NCs were composed of two different materials: the starting $Cu_{2-x}Te$ and rs-PbTe or SnTe (see Figure 6c and Table S1). As evidenced by HRTEM images, each NC is composed of a $Cu_{2-x}Te$ core and a polycrystalline PbTe or SnTe shell. In both cases, the high deviation of lattice constants from that of $Cu_{2-x}Te$ (the (200) planes are at 3.2 Å and 3.1 Å in PbTe and SnTe, respectively) does not allow for the formation of a single domain shell characterized by low indices planes (the strain at the interface is too high). Instead, the shell was composed of multiple domains, some of them evidencing planes with high Miller indices. Also, some of these domains were partially amorphous, especially for the $Cu_{2-x}Te \rightarrow SnTe$ case.

**Figure 6.** HRTEM images of $Cu_{2-x}Te@PbTe$ (a) and $Cu_{2-x}Te@SnTe$ (b) core@shell heterostructures obtained by partial CE of $Cu_{2-x}Te$ NCs with $Pb^{2+}$ and $Sn^{2+}$ cations, respectively. The insets are HRTEM images at higher magnifications showing in details the border (green dashed line) between the $Cu_{2-x}Te$ core and the (a) PbTe or (b) SnTe shell. As evidenced by HRTEM images, both CE processes lead to a polycrystalline shell material. (c) XRD patterns obtained from dropcast solutions of $Cu_{2-x}Te@PbTe$ (red pattern) and $Cu_{2-x}Te@SnTe$ (blue pattern) heterostructures with the corresponding bulk reflections: SnTe (dark blue bars, ICSD number 98-065-2759) and altaite PbTe (dark red bars, ICSD number 98-006-3099). For a more clear comparison the experimental reflections measured for the starting $Cu_{2-x}Te$ NCs (see also Figure 1) were also reported by means of grey bars.

Interestingly, previous works demonstrated that some reported core@shell heterostructures formed by CE are metastable (kinetically accessed) and evolve into thermodynamically stable Janus-like structures upon annealing in vacuum or e-beam irradiation.[33-52] In order to assess the stability of our systems, annealing experiments were performed by heating $Cu_{2-x}Te@SnTe$ and $Cu_{2-x}Te@PbTe$ NCs in the TEM. As shown in Figure 7, the morphology of both NCs evolved from core@shell to Janus-like upon heating, while retaining the overall size and morphology. It appears that, in both cases, the overall interface energy was minimized by simply reducing the interfacial area between the two structures. These experiments confirm the metastable nature of our core@shell structures, whose appearance, here, is most likely ascribable to the low diffusion rate of octahedrally coordinated ions that, presumably, are able to rapidly access the host NCs surface (thanks to the high density of Cu vacancies), where they accumulate since their inward diffusion is slow. Thus the nucleation of the product SnTe or PbTe phase is limited to the host NCs surface, where it takes place unselectively.

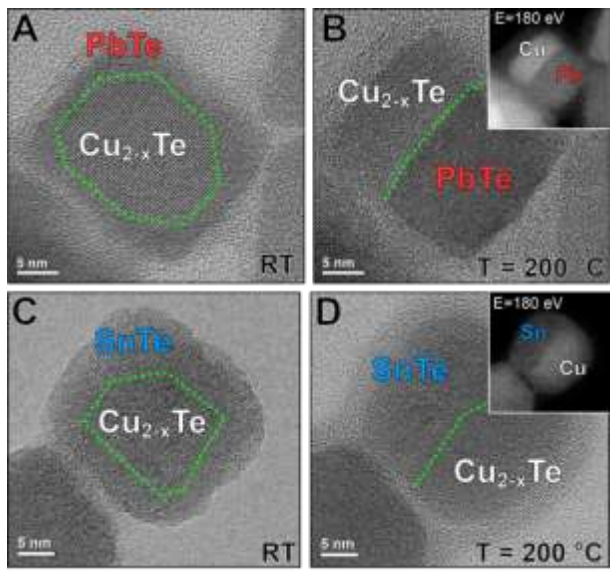

**Figure 7.** HRTEM images of $Cu_{2-x}Te@PbTe$ and $Cu_{2-x}Te@SnTe$ core@shell NCs before (a and c, respectively) and after (b and d, respectively) the *in situ* annealing treatment at 200 °C. In both cases the annealing causes the transition from a core@shell architecture to a more stable Janus-like morphology, as a consequence of the minimization of the interfacial energy (b-d). The insets in b and d show filtered images at 180 eV, used to identify unambiguously the $Cu_{2-x}Te$ domain in the particles.

The formation of $Cu_{2-x}Te@SnTe$ core@shell architectures, observed in this work, is somehow surprising when considering that the corresponding copper selenide NCs, that is $Cu_{2-x}Se$, form Janus-like $Cu_{2-x}Se/SnSe$ NCs upon $Sn^{2+}$ CE at 100 °C, as we previously reported in a recent work.[47] Given the similarities between $Cu_{2-x}Te$ and $Cu_{2-x}Se$ NCs and considering that $Sn^{2+}$ ions opt for an octahedral coordination with both $Te^{2-}$ and $Se^{2-}$ anions, in fact, similar morphologies of the resulting NCs were expected upon partial CE. The formation of Janus-like $Cu_{2-x}Se/SnSe$ architectures in our previous work[47] suggests that the ion diffusion rate in that system, if compared to the $Cu_{2-x}Te \rightarrow SnTe$ case, is much higher, favoring the creation of a single SnSe domain per NC. This difference was tentatively attributed to the higher reaction temperature at which the $Cu_{2-x}Se \rightarrow SnSe$ reaction was performed (100 °C as opposed to RT of the present reactions on $Cu_{2-x}Te$ NCs). In order to test our hypothesis we performed a control experiment in which we exposed $Cu_{2-x}Se$ NCs to $Sn^{2+}$ cations at RT, using the same synthetic conditions as for $Cu_{2-x}Te$ NCs of this work (see SI for further information). As evidenced by XRD and HRTEM analysis, the product of this experiment consisted of heterostructures having a central domain made of $Cu_{2-x}Se$ surrounded by multiple SnSe crystal domains, resembling a core@partial-shell morphology (see Figure S7 and S8).

In the light of these results, we can rationalize the formation of core@shell or segmented heterostructures upon partial CE as the outcome of two main different competing processes: the nucleation of the product material in/on the starting NC and the cations interdiffusion inside the host lattice. The nucleation rate of the product phase is determined by factors such as the rate at which guest cations can access the NC's surface and the speed at which host cations are replaced and, thus, solvated by ligands.[53] The energetics and the kinetics of these processes can be, indeed, very different for each specific facet of the host NC as diverse surface energies can be involved. The nucleation process is, indeed, favored by the presence of surface cation vacancies because, especially at the early stage of the process, they provide empty "sites" for guest ions to access the NC.[10] The ion diffusion rate, on the other hand, is influenced by the possibility of cations to diffuse *via* vacancies and/or interstitial sites. The type of coordination of the exchanging cations has a profound impact on their diffusion rates, as it dictates their effective ionic radii and through which sites they can move. Similarly to the nucleation process, the diffusion rate of the exchanging species is strongly dependent on the reaction temperature and on the presence of cation vacancies in the host NCs.[11-12,15,38] When the unselective nucleation of the product material on the host NC surface is faster than the cations interdiffusion process, kinetically accessed core@shell heterostructures can form. Conversely, if the cation diffusion rate is faster than the nucleation rate, the new phase can form in a preferred region(s) of the host NC, and then it grows from there at the expenses of the starting material. This process leads to segmented or Janus-like heterostructures, in which both interfacial energy and strain are usually minimized. There exist also peculiar cases in which the host and/or guest ion diffusion is slow and the nucleation of the product material is energetically allowed only on specific facets of the host NC. Under these conditions, segmented or more elaborate heterostructures form upon partial CE.[35,38,53-55] These cases typically involve NCs with a high aspect ratio, like nanorods or nanodiscs, where some facets are much more stable than others. The tips of a nanorod or the edges of a nanoplate are, indeed, the most reactive sites where the product phase of a CE reaction initially forms.

## CONCLUSION

In conclusion, we have shown that cubic-shaped $Cu_{2-x}Te$ nanocrystals can be used as templates to access metal telluride nanostructures by cation exchange at room temperature. Trioctylphosphine alone, in some experiments, could not drive the complete exchange and the concomitant use of a diamine (N,N-dimethylethylenediamine) was shown to improve the extraction of the copper ions, allowing the CE transformation. The coordination number of the entering cations was found to have a profound influence on the kinetics of the exchange reaction. The occurrence of a specific morphology in partial cation exchange experiments was explained as a result of two competing processes: the nucleation of the product material and the ion diffusion. In our $Cu_{2-x}Te$ nanocrystals, characterized by a high density of Cu vacancies, indeed, the nucleation rate of the product materials should be the same on all the facets of the nanocubes. On the other hand, the diffusion rate of the entering cations strongly depends on their type of coordination with the lattice. The tetrahedrally coordinated cations can rapidly diffuse into the host nanocrystals, with the consequent formation of Janus-like architectures. Octahedrally coordinated ions, being not able to exploit all the vacant sites in the $Cu_{2-x}Te$ structure, have a limited diffusion rate which leads, upon partial cation exchange, to the formation of core@shell architectures. These structures, if annealed under vacuum in TEM, evolve into more stable Janus-like heterostructures at relatively low temperatures. We also showed that the temperature at which such exchange reactions are performed can be of outmost importance in selectively tuning the ion diffusion rate and thus in tailoring the morphology of the resulting heterostructures.

We believe that our findings can be of help in guiding the design of other types of heterostructures. Starting from the conclusions of this work one can study, for example, if raising (or lowering) the temperature of a certain CE reaction might help in varying the ion diffusion rate in order to make it faster (or slower) than the nucleation rate of the product material. Indeed, a fine tuning of these rates, as we showed in our manuscript, could enable the control over the morphology of the resulting heterostructures. On the other hand, the generalization of our findings to other (nanostructured) materials is not always straightforward as each system requires a specific analysis. Ion diffusion in solids depends indeed on many parameters, and it is faster in "open" crystalline structures (as the fast ionic conductors classes of materials, to which copper chalcogenides belong) than in close-packed structures.[56] In the latter case for example, the exchange might proceed preferentially on the surface of the particles, regardless of the type of coordination of the entering cations, and partial CE could result always in core@shell architectures. Other parameters to be taken into account are shape and crystal lattice anisotropies, as already mentioned in the previous section. Particles with anisotropic crystal structure and exposing highly reactive facets/sites (for example rods or discs) would tend to initiate cation exchange reactions on these reactive regions, thus resulting in segmented architectures as intermediate exchange products, again regardless of the type of cation coordination.

## ASSOCIATED CONTENT

**Supporting Information**. Details on size distribution analysis of $Cu_{2-x}Te$ NCs, low resolution TEM images of NCs obtained after full and partial CE experiments, annealing experiments on HgTe and CdTe NCs, CE reactions between $Cu_{2-x}Se$ NCs and $Sn^{2+}$ ions and the $Cu_{2-x}Te$ XRD pattern indexing. This material is available free of charge via the Internet at http://pubs.acs.org.

## AUTHOR INFORMATION

### Corresponding Author

luca.detrizio@iit.it, liberato.manna@iit.it

## ACKNOWLEDGMENT

We acknowledge funding from the European Union under grant agreements n. 614897 (ERC Grant TRANS-NANO).

Insert Table of Contents artwork here

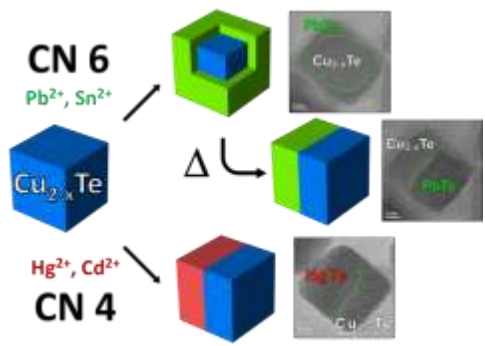